\documentclass[useAMS,usenatbib]{mnras}
\usepackage{amssymb}
\usepackage{amsmath}
\usepackage{txfonts}
\usepackage{graphicx}
\usepackage{subcaption} 
\usepackage{epsfig}
\usepackage{xspace}
\usepackage{url}
\usepackage{comment}
\usepackage[colorinlistoftodos]{todonotes}

\usepackage{natbib}
\def\swift{\textit{Swift}\xspace}

\def\xrt{\textit{Swift}/XRT\xspace}

\def\lat{\textit{Fermi}/LAT\xspace}

\def\psrb{PSR~B1259$-$63\xspace}

\title[2024 periastron passage of PSR B1259$-$63]{Multiwavelength coverage of the 2024 periastron passage of PSR~B1259$-$63 / LS~2883}
\author[Chernyakova et al.]{\parbox{\textwidth}
{
    M. Chernyakova,$^{1,2}$\thanks{E-mail: \href{mailto:masha.chernyakova@dcu.ie}{masha.chernyakova@dcu.ie}} 
    D. Malyshev,$^{3}$
    B. van Soelen,$^{4}$
    A. Finn Gallagher,$^{1}$
    N. Matchett,$^{4}$
    T.~D. Russell,$^5$\\
    J. van den Eijnden,$^{6}$
    M.~E. Lower,$^{7}$
    S. Johnston,$^{7}$
    S. Tsygankov,$^{8}$
    A. Salganik,$^{8}$
    Iu. Shebalkova$^{1}$
}
\\ \\ 
$^{1}$School of Physical Sciences and Centre for Astrophysics \& Relativity, Dublin City University, Glasnevin, D09 W6Y4, Ireland.\\
$^{2}$Dublin Institute for Advanced Studies, 31 Fitzwilliam Place, Dublin 2; \\
$^{3}$Institut f{\"u}r Astronomie und Astrophysik T{\"u}bingen, Universit{\"a}t T{\"u}bingen, Sand 1, D-72076 T{\"u}bingen, Germany\\ 
$^{4}$Department of Physics, University of the Free State, PO Box 339, Bloemfontein 9300, South Africa \\
$^{5}$ INAF - IASF Palermo, via Ugo La Malfa, 153, I-90146 Palermo, Italy\\
$^{6}$ Department of Physics, University of Warwick, Coventry CV4 7AL, UK \\
$^{7}$Australia Telescope National Facility, CSIRO, Space and Astronomy, PO Box 76, Epping, NSW 1710, Australia\\
$^{8}$Department of Physics and Astronomy, FI-20014 University of Turku, Finland\\
}

\date{Accepted XXXX. Received YYYY; in original form ZZZZ}

\begin{document}
\label{firstpage}
\pagerange{\pageref{firstpage}--\pageref{lastpage}}
\maketitle

\begin{abstract}
\psrb is a gamma-ray binary system with a 48~ms radio pulsar orbiting around an O9.5Ve star, LS 2883, in a highly eccentric $\sim3.4$~yr long orbit. Close to the periastron the system is detected  from radio up to the TeV energies due to the interaction of the stellar wind from LS~2883 and the pulsar's relativistic outflow. Observations of the last four periastron passages,  taken in 2010-2021, demonstrate periastron-to-periastron variability at all wavelengths, probably linked to the state of the Be star's decretion disk.
In this paper we present the results of our optical, radio and X-ray observational campaigns on \psrb performed in 2024 accompanied with the analysis of the publicly available GeV \lat data. We show that this periastron passage was characterised by the early flaring of X-rays before the periastron passage and GeV emission after the periastron passage, which can be explained by a larger size of the decretion disk as supported by the optical observations. The structure of the GeV flare is also in agreement with the disruption of the large dense disk.  The possible X-ray/radio correlation was observed only during the post-periastron rise of X-ray and radio emission. 
\end{abstract}


\section{Introduction}
First discovered  during the Parkes   high-frequency radio survey in 1990 \citep{1992MNRAS.255..401J}, \psrb is a classical gamma-ray loud binary. In this system a young 48 ms radio pulsar is orbiting around an O9.5Ve massive star, LS~2883,  in a highly eccentric ($e = 0.87$) 3.4~years long orbit \citep{2011ApJ...732L..11N,shannon14,PSRB1259-2018_distance}. The decretion disk of LS~2883 is inclined to the orbital plane  by about 35$^\circ$ \citep{shannon14} and the  pulsed radio emission is eclipsed by the disk around periastron. Unpulsed radio emission appears at about the time the pulsar approaches the disk for the  first time, showing two maxima (peaks) during the first and second disk crossing,  which greatly exceed the flux level of the pulsed emission.
The position of these radio peaks roughly coincides with the peaks in the X-ray light curve \citep[see e.g.][]{Chernyakova21_psrb}. 

The observed radio to TeV emission from this system is generally believed to be the result of the interaction of the relativistic electrons of the pulsar wind with the LS~2883 radiation and wind outflow. The radio and X-ray emission is usually assumed to be a result of synchrotron emission \citep{tavani97, 2002MNRAS.336.1201C, Chernyakova20_psrb}, though note that the emission region(s) and electron population(s) are still under debate. The TeV emission is due to the Inverse Compton (IC) scattering of the soft photons on the relativistic electrons of the pulsar wind \citep{Kirk99,Chernyakova20_psrb}. 
The origin of the GeV emission from this system is still under debate. The observed variability of the GeV emission on timescales as short as 15 minutes \citep{FERMI_PSRB2018} and the lack of correlation with any other energy band excludes one-zone models \citep[e.g.][]{2011ApJ...736L..10T,kong12} and  makes it difficult to attribute the observed GeV emission to either synchrotron or IC emission \citep[e.g.][]{Khan12,dubus_cerutti13,2015ApJ...798L..26T,Chen2019}.   
Optical observations show an increase in the equivalent width of the H\,$\alpha$ emission line around periastron, and a subsequent decrease starting around 30 days after periastron, which is linked to changes in the structure of the circumstellar disk \citep{Chernyakova2017,Chernyakova21_psrb,Chernyakova24, 2016MNRAS.455.3674V}. 
A possible solution to the origin of the GeV flare was proposed by \cite{Chernyakova20_psrb}. In this paper it was shown that a combination of IC emission with the bremsstrahlung emission of the relativistic electrons drifting along the shock and interacting with the clumps produced due to the disruption of the disk can explain both the observed luminosity and the structure of the flare.

 Observations of \psrb  before the 2021 periastron allowed us to propose a relation between the GeV flare and the disruption of the disk, suggested by the decrease of the equivalent width of the H\,$\alpha$ emission line \citep{Chernyakova20_psrb}. Observations of \psrb during its 2021 periastron passage, however, have clearly demonstrated that the evolution of the disk is not directly linked to the start of the GeV flare, as the evolution of the H$\alpha$ equivalent width in 2014 and 2021 was very similar, while the start of GeV flare in 2021 was very delayed and began only $\sim$60 days after periastron \citep{Chernyakova21_psrb}. Observations of the 2021 periastron passage also demonstrated other new features, like the correlation of the radio/X-ray emission during the second X-ray peak (and the subsequent disappearance of this correlation afterwards), the sudden appearance of the third X-ray peak \citep{Chernyakova21_psrb,Chernyakova24}, and evidence for X-ray/TeV correlation \citep{HESS_psrb24}.

The  2024 periastron passage of \psrb occurred on June 30,  $t_p=$MJD 60491.592 \citep{shannon14}.
In order to better understand the physical processes in this interesting system, we undertook an intensive radio, optical and X-ray monitoring campaign before, during, and after the 2024 periastron.
In this paper, we report on the initial results of this campaign. In Section 2 we describe the details of radio, optical, X-ray and GeV data analysis. The results of the observations are presented and discussed in Section 3.

\section{Data Analysis}
\subsection{Radio Data}
For the 2024 periastron, \psrb was observed in the radio band with ATCA, under project code C3582. This observing campaign began approximately 20 days before periastron (2024 June 11), finishing 79 days after the periastron (2024 Sep 17). 
Data was recorded at two central frequencies, centred at 5.5\,GHz and 9\,GHz, with 2\,GHz of bandwidth at each frequency \citep{wilson2011}.

The campaign consisted of 31 observations in total. Observations typically lasted for an average of $\sim$4-hours. During each observation, two calibrators were also observed (in addition to \psrb). 
Bandpass and flux density calibration was carried out with either PKS B1934$-$638 or PKS B0823$-$500, depending on source visibility, where PKS B1934$-$638 was the preferred calibrator.
We also used the nearby unresolved bright radio source PKS J1322$-$6532 for phase calibration. 

The Common Astronomy Software Applications ({\sc CASA}), version 6.6.3.22 \citep{2022CASA} was used to analyse the data. When starting the analysis, the data had to be flagged by removing radio frequency interference (RFI) and systematic issues.  
The data were then calibrated following standard procedures in {\sc CASA}.\footnote{\url{https://casaguides.nrao.edu/index.php/Main_Page}} Imaging was carried out using \texttt{tclean} within {\sc CASA}. The calibrated integrated flux value was found by fitting a 2-D Gaussian to the point source in the image with the \texttt{imfit} function, fixing the Gaussian full-width-at-half-maximum values to those of the synthesized beam.

\subsection{Optical Data}

\begin{figure}
    \centering
    \includegraphics[width=\linewidth]{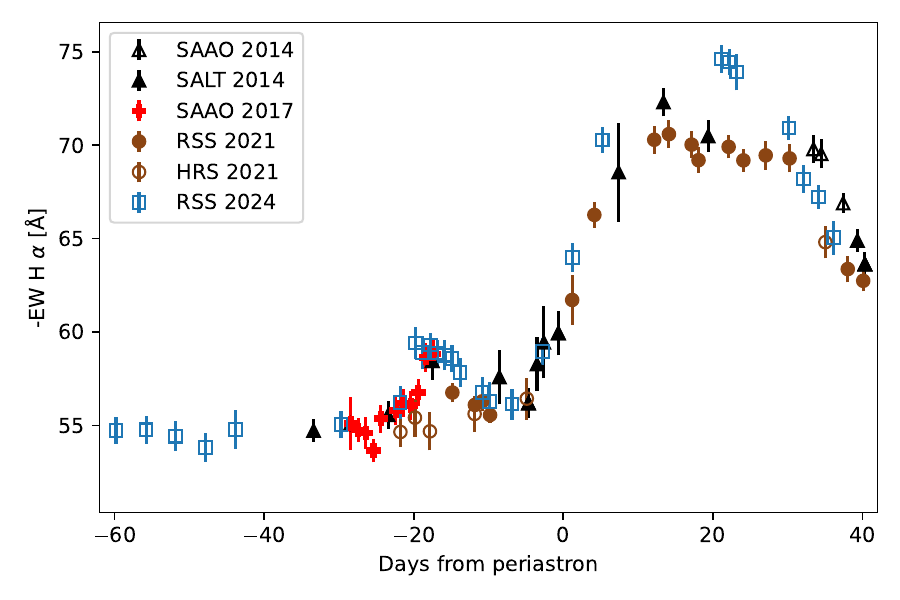}
    \caption{Equivalent width (EW) of the H\,$\alpha$ emission line around the 2024 periastron passage of \psrb, compared to the previous periastra.}
    \label{fig:salt}
\end{figure}

Optical coverage of the 2024 periastron passage of PSR B1259$-$63 was carried out with the Robert Stobie Spectrograph \citep[RSS;][]{burgh03} on the Southern African Large Telescope \citep[SALT;][]{buckley06}. Observations were obtained from $\approx$60\,d before (2024 May 1) until $\approx36$\,d after periastron (2024 Aug 5) in order to trace the behaviour of the H\,$\alpha$ emission line (produced by the circumstellar disk) as the pulsar passed through periastron. All observations were taken with the PG2300 grating, with the 0.6\arcsec\ long slit (achieving a resolving power of $R\approx$11\,000), and consisted of 15$\times$30\,s exposures. The data reduction and spectral extraction followed the same procedure as discussed in \citet{Chernyakova21_psrb, Chernyakova24}.   All the data was first pre-reduced by the SALT pipeline, and then flat correction, wavelength calibration and spectral extraction was performed, and each night's observations were averaged together, continuum corrected with a low-order polynomial, and barycentric corrected, using a {\sc python} script relying on {\sc numpy, scipy, astropy} and {\sc pyraf}. The equivalent width (EW) of the H$\alpha$ emission line was then measured by numerically integrating over the wavelength range $\lambda= 6528 - 6607$\,\AA{}, and the uncertainty was estimated following  \citet{vollmann06}, with the signal-to-noise ratio estimated as in \citet{stoehr08}. The variation in H\,$\alpha$ EW for the 2014 - 2024 periastra is shown in Fig.~\ref{fig:salt}.

\subsection{X-ray Data}

\subsubsection{Swift}

\begin{figure*}
    \centering
    \includegraphics[width=\linewidth]{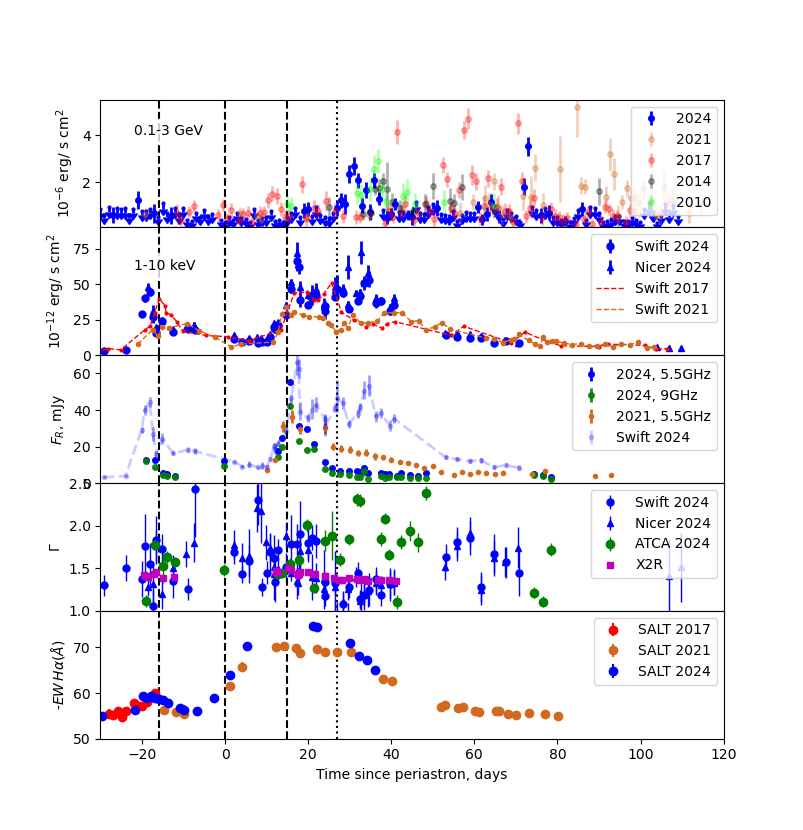}
    \caption{The \psrb light curves from radio to GeV energies. The dashed lines correspond to periastron and to the moments of the disappearance (first non-detection, $t=t_p-16$) and the reappearance (first detection, $t=t_p+15$) of the pulsed emission, as observed in 2010 \citep{2011ApJ...736L..11A}.
    The dotted line corresponds to the first appearance of the detection in GeV band on a day time scale at the beginning of the 2024 GeV flare. All historical data in this figure are taken from \citep[][and references therein]{Chernyakova21_psrb, Chernyakova24}.
 \textit{Panel a:} \lat flux measurements in the E $>$ 100 MeV energy range with a daily bin size.
\textit{Panel b:} absorbed 1-10 keV X-ray flux.
     \textit{Panel c:} radio flux densities in mJy. Scaled 2024 X-ray Swift data is also shown in this panel with a blue dashed line for comparison. 
       \textit{Panel d:}  Evolution of the X-ray (blue) and radio (green, $1- \alpha$) photon indexes around periastron. In magenta we show the photon indexes that would explain the overall radio to X-ray spectrum   {(using data points separated by no more than 1.5~days)}. 
     \textit{Panel e:} H\,$\alpha$ equivalent width.
     }
    \label{fig:multiwavelength_lc}
\end{figure*}

The 2024 periastron passage of \psrb was closely monitored by the Neil Gherels Swift Observatory \citep[\swift;][]{2004ApJ...611.1005G}. We have analyzed all available data taken from June 1 to September 9, 2024. The~data was reprocessed and analysed as suggested by the \xrt team\footnote{See \xrt \href{https://swift.gsfc.nasa.gov/analysis/xrt_swguide_v1_2.pdf}{data analysis guide}}  with the \texttt{xrtpipeline v.0.13.7} and \texttt{heasoft v.6.34} software package. The spectral analysis of \xrt spectra was performed with \texttt{XSPEC v.12.14.1}. The spectrum was extracted from a circle of radius $36''$ around the position of \psrb, and~the background estimated from a co-centred annulus with inner/outer radii of $60''/300''$. We performed the fit of the spectrum, grouped to have at least 1 photon per energy bin, using ``cstat'' statistic~\citep{cash79, Wachter1979} by an absorbed power-law model (\texttt{cflux*tbabs*po}) in 0.3--10~keV range. The flux of the source, hydrogen column density and the slope of the power-law were treated as free parameters during the fit. The~uncertainties of the \xrt fluxes shown in Figure~\ref{fig:multiwavelength_lc} are $1\sigma$ confidence intervals.

\subsubsection{NICER}
NICER is an externally attached payload on the International Space Station, delivered in 2017 \citep{2016SPIE.9905E..1HG}. The primary instrument of NICER is the X-ray Timing Instrument (XTI). It consists of 56 pairs of X-ray concentrators (XRC) and silicon drift detectors (SDD). Each SDD detects individual incoming photons collected by the XRC. NICER/XTI operates in the energy range of $0.2–12$ keV, with a sufficiently large effective area of 1700 cm$^{2}$ at 1 keV, high sensitivity, and an unprecedented precision in photon arrival time measurements -- less than 300 ns. However, the design of NICER does not allow it to produce images, meaning that the photons are registered from a region of the sky that spans several arcmin.

We used data from the monitoring campaign of \psrb, where spectral extraction was possible and the spectrum was not dominated by the background noise, consisting of 27 observations covering MJD 60472--60539 (ObsIDs 5203610101-6, 8-15, 17-21, 27, 31-32, 35-36).
NICER data were reduced using the {\sc nicerdas} software and {CALDB} version 20240206 according to the official data analysis threads.\footnote{\url{https://heasarc.gsfc.nasa.gov/docs/nicer/analysis_threads/}} The data were processed using the \texttt{nicerl2} task, then the cleaned event files were used to generate spectral products using the \texttt{nicerl3-spect} task, incorporating the SCORPEON background model. For each observation an ``orbit-night" spectrum was extracted using standard threshold settings. If a ``night" spectrum had zero effective exposure, a ``day" spectrum was extracted instead. NICER spectra were fitted using a simple absorbed power-law. In accordance with the official SCORPEON analysis manual, noise\_norm, lhb\_em and gal\_nh parameters of SCORPEON model were frozen.\footnote{\url{https://heasarc.gsfc.nasa.gov/docs/nicer/analysis_threads/scorpeon-xspec/}} The \psrb spectra were approximated using simple absorbed power-law using ``cstat'' statistic.

\subsection{GeV Data}

\begin{table*}
  \caption{The best-fit model parameters for \lat observations of \psrb for different periods around the 2024 periastron. The Model column indicates the model used to fit the spectrum. The super-exponential cutoff-powerlaw (ECSPL) model is defined as $dN/dE = N_0 (E/\mbox{1 GeV})^\Gamma \exp(E/E_c)^\beta$. The powerlaw (PL) model corresponds to a ECSPL model with $E_c=\infty$. The TS column summarizes the test-statistics of the detection of \psrb during the corresponding period. The parameters denoted by $^\dagger$ were kept fixed during the fit due to a low detection significance of the source during this period.}
    \centering
    \begin{tabular}{|c|c|c|c|c|c|c|}
    \hline
     Period    &  Model & Norm $N_0$        & Index, $\Gamma$         & Cutoff, $E_c$     & $\beta$       & TS \\ \
     days      &        &$10^{-9}$~ph/cm$^2$/s&               &    MeV     &               &    \\
     \hline
      -20 to 0 &  PL    &   $(0.7\pm 0.4)\cdot 10^{-2} $                  & $2.7\pm 0.3$  &   --       &  --           &  17 \\
      0 to +20 &  PL    &   $(0.4\pm 0.2)\cdot 10^{-2} $                 & $2.7^\dagger$ &   --       &  --           &  6.5 \\
    +19 to +77 &  ECSPL & $1.75\pm 0.7$       & $1.5\pm 0.2$  &  $42\pm 8$ & $0.49\pm 0.04$&  690 \\
    \hline
    \end{tabular}
  
    \label{tab:fermi_bestfits}
\end{table*}

We performed an analysis of \lat data with Fermitools (v. 2.2.0 released 21 June 2022) using the latest Pass 8 reprocessed data (P8R3) from the SOURCE event class. 
The binned likelihood analysis was performed for photons within the energy range 0.1--300~GeV that arrived between MJD 60401 and 60601 ($-90$ to $+110$~days around the periastron) within a $20^\circ$-radius region around \psrb's position. The~selected maximum zenith angle was $90^\circ$. The~performed analysis relies on the fitting of the spectral and spatial model of the region to the observed data in a series of energy and time~bins. 

The adapted model of the region included templates for the Galactic and isotropic diffuse emission components provided by the \lat collaboration, as~well as all sources from the latest \lat 4FGL-DR4 catalogue~\citep{4FGL}, with~the spectral templates selected according to the catalogue. At~the initial stage of the analysis we assumed all spectral parameters of sources within $15^\circ$ around \psrb to be free parameters and fixed all spectral parameters of sources between $15^\circ$ and $20^\circ$ to their catalogue values. We performed the fitting of the described model to all the available data. For the subsequent analysis, we fixed all (except normalisations) free spectral parameters to their best-fit values, and~removed all of the weak sources detected with test-statistics $TS<1$ from the~model.

For the next step of the analysis, we split the 0.1--3~GeV data over a series of 1-day and 1-week bins aiming to produce the light curve of \psrb in the corresponding energy band (see Figure~\ref{fig:fermi_lc}, left panel). All upper limits presented were calculated using the \texttt{IntegralUpperLimits} module included in Fermitools for $TS<4$ cases and correspond to a 95\% confidence~level. Aiming to study short-timescale variability of the system we also considered time bins of a variable duration. To achieve this, we defined time bin durations to accumulate 9 photons in the energy interval 0.1-3~GeV in a $1^\circ$-radius region around \psrb, similar to the approach used in~\citet{Chernyakova21_psrb} to detect  short-timescale variability during the 2021 periastron passage. Alternatively, we considered ``Bayesian blocks'' time binning,\footnote{Using \href{https://docs.astropy.org/en/stable/api/astropy.stats.bayesian_blocks.html}{astropy implementation} of \citet{bayesian_blocks} algorithm with a false-alarm prior probability $p_0=0.05$} i.e. time bins during which the count rate within 0.1-3~GeV in a $1^\circ$-radius region around \psrb position is approximately constant. We note a good overall agreement between the adapted approaches (see corresponding light curves in Fig.~\ref{fig:fermi_lc}, right panel).

The obtained light curves suggest the enhancement of the GeV emission during +19 to +77~days (MJD 60510 -- 60568) after periastron. Below we refer to this period as the ``\lat flare'' during the 2024 periastron passage. To study the potential spectral-shape variability of the source during different periods of the periastron passage we built the spectrum of \psrb for the periods $(-20;0)$, $(0; +20)$ days around periastron, as well as during the \lat flare period. These periods were selected to be similar to those used by \citet{Chernyakova21_psrb} which allows us to make a direct comparison between the current and 2021 periastron passages. The spectra of the 2024 and 2021 periastron passages are shown in Figure~\ref{fig:fermi_spectra}. 

We found that \psrb is detected during all considered time periods with a test-statistic (TS) of $TS\sim 17$ for $(-20;0)$~days; $TS\sim 6.5$ for $(0; +20)$~days and $TS\sim 690$ for the period of the flare ($+19 to +77$~days). Due to a low detection significance of the source during $(0; +20)$~days and $(-20;0)$~days periods we fit the spectra during these periods with a powerlaw (PL) model, additionally fixing the powerlaw index for the $(0; +20)$~days period to $2.7$. For the flaring period the spectrum of \psrb is significantly curved and can be described with a superexponential cutoff-powerlaw (ECSPL) model. The best-fit spectral parameters in the $0.08-2$~GeV band are summarized in 
Table~\ref{tab:fermi_bestfits}.

\section{Results and Discussion}

\begin{figure*}
    \centering
    \includegraphics[width=0.48\linewidth]{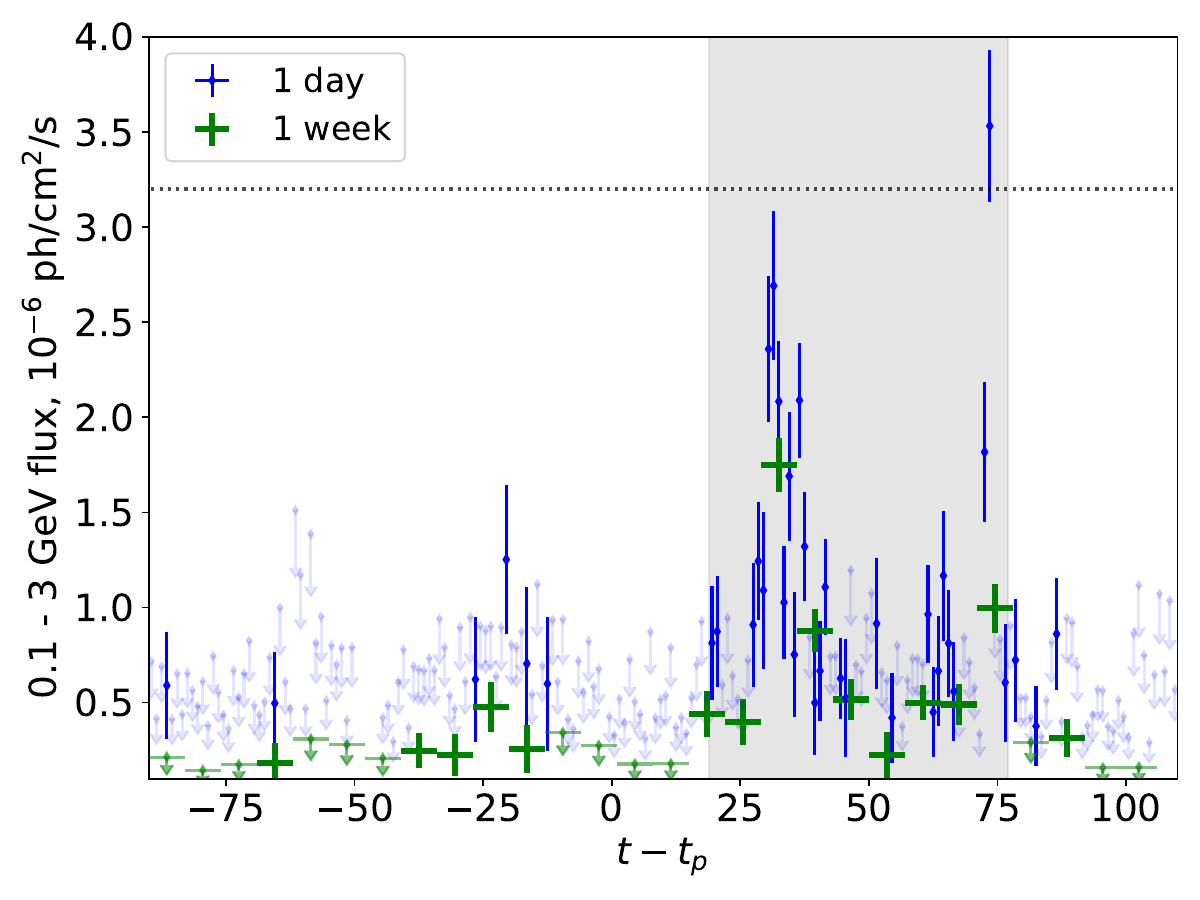}
    \includegraphics[width=0.48\linewidth]{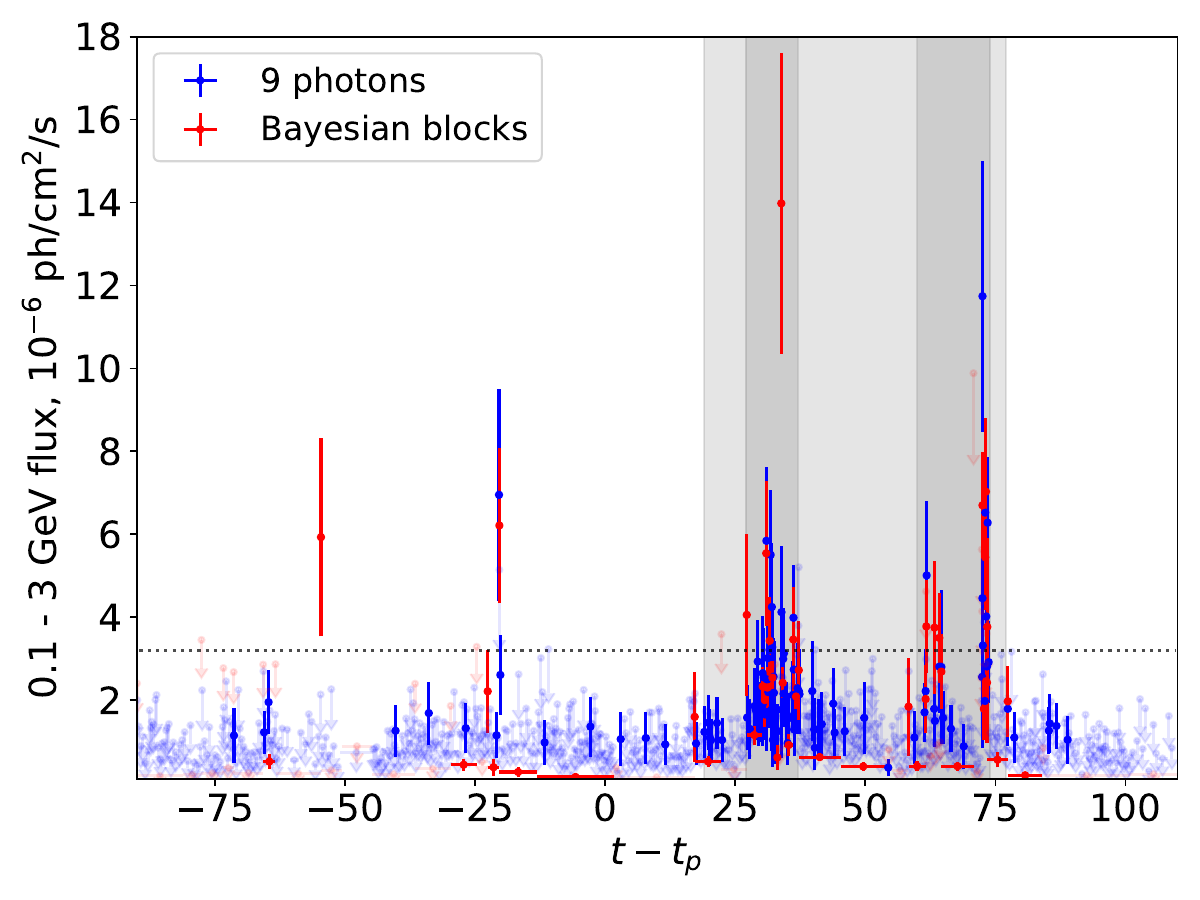}
    \caption{\lat light curve of \psrb in the 0.1-3~GeV energy band. The left panel shows daily and weekly time bins. The right panel presents time-variable bins -- bins accumulating 9 photons in $1^\circ$-radius region around \psrb (blue points) and bins of a constant count-rate in $1^\circ$-radius circle \citep[``Bayesian blocks''][red points]{bayesian_blocks}; see text for details. The black dotted line illustrates the spin-down luminosity flux for \psrb. The light gray region shows the period $+19-+77$~days used to derive the spectrum of \lat flare, see Fig.~\ref{fig:fermi_spectra}. Dark-grey regions on the right panel additionally show the period of enhanced short-timescale variability ($+27$ to $+37$~d and $+60$ to $+74$~d). }
    \label{fig:fermi_lc}
\end{figure*}

\begin{figure}
    \centering
    \includegraphics[width=\linewidth]{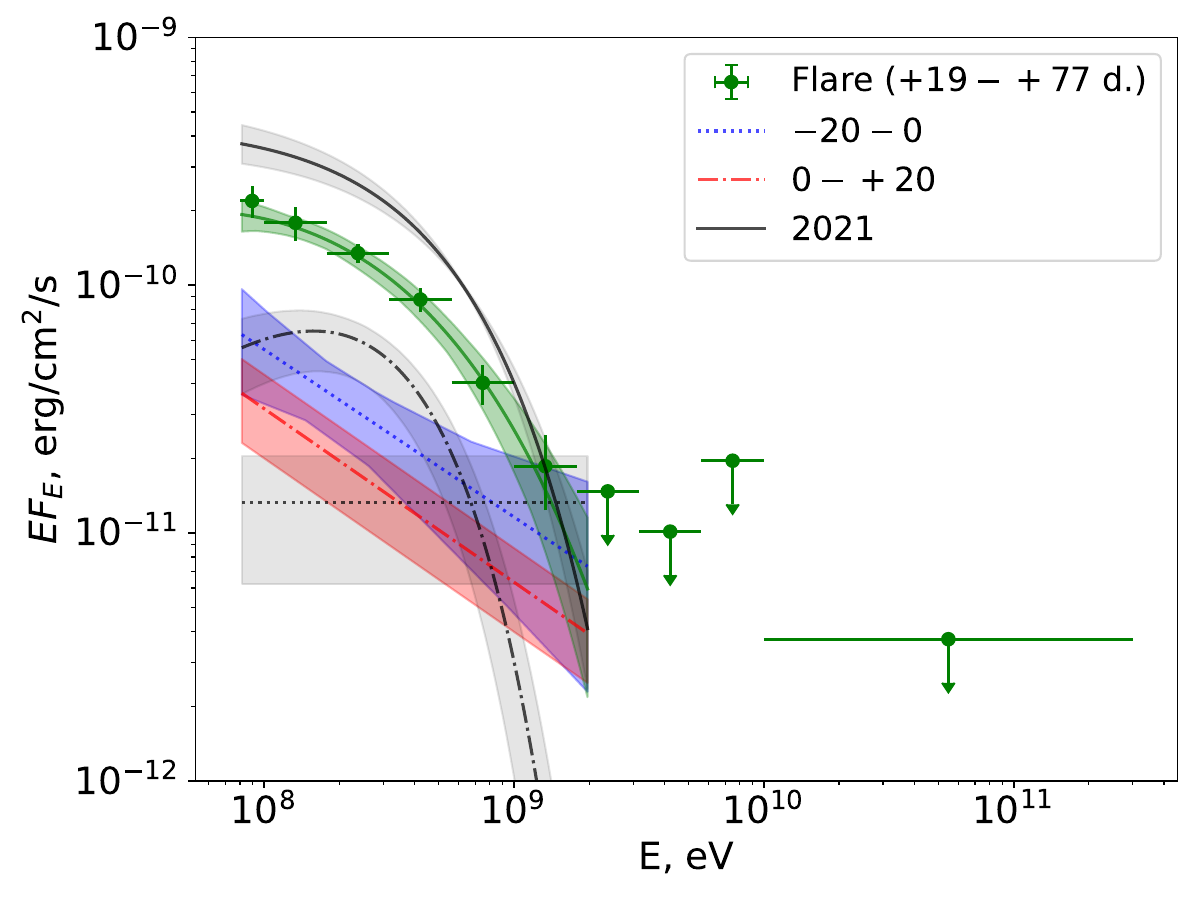}
    \caption{
    Best-fits of \lat spectra of \psrb before ($-20$ to $0$ days, blue dotted line) and after ($0$ to $+20$ days, red dot-dashed line) the periastron passage with a powerlaw model. The green solid line illustrates the best-fit of the spectrum  during the period of the flare ($+19 to +77$~days). Green points show the \lat spectrum during this period. Same line style black lines show the results from 2021 periastron passage \citep[adapted from][]{Chernyakova21_psrb}). Shaded regions correspond to $1\sigma$ uncertainty regions around the best-fit models.  }
    \label{fig:fermi_spectra}
\end{figure}

The 2024 periastron passage of \psrb demonstrated a peculiar behaviour of the system from radio to GeV energies.
The H$\alpha$ equivalent width (Figs.~\ref{fig:salt} \& \ref{fig:multiwavelength_lc}) showed a  general behaviour which was similar to previous periastra: the EW was fairly constant before the time of the first disk crossing; it increased and showed a local maximum around the first disk crossing before decreasing; the EW then increased and peaked after periastron, before decreasing again. However, there are notable differences this time. First, the increase around the  first disk crossing was noticeably larger than it was  in 2021, and second the first maximum in 2024 appears to be $\sim 1$\,day earlier than in 2017. Unfortunately, the gaps in the data do not allow us to place a tighter constraint on this. Finally, the peak of  EW after periastron is significantly larger than was ever observed before.

This difference in the EW would be inline with the pulsar interacting earlier with the circumstellar disc, either because of a larger disc, or due to a small shift in the disk orientation. The presence of the large disk up to $t_p+29$ is also indicated by infrared observations~\citep{alma2024}.

Such an assumption is further supported by the X-ray observations, see Fig.~\ref{fig:multiwavelength_lc}. 
The  bright and narrow first X-ray maximum was observed at $\sim -18$~days, i.e., a few days earlier than ever observed before.  
The second X-ray peak is characterized by a sharp rise at $\sim +15$~days, in line with previous observations, followed by a period of highly variable enhanced emission until at least $+40$~days. Such a behavior can be understood  if, during the second passage, the pulsar perturbed the  disk and/or dragged a large amount of the decretion disk's material along with it. In this case the apex of the pulsar-stellar wind interaction cone could have remained relatively close to the pulsar for a longer time, i.e. in a higher  magnetic field region resulting in enhanced X-ray emission. The strong quasi-random X-ray variability at $+20$ to $+40$~days can be connected to large-scale density variations in the perturbed disk which push the tip of the emission cone closer to/further from the pulsar.

Unfortunately our ATCA data does not cover the rise in X-ray emission during the first X-ray peak which, therefore, does not not allow us to study the radio/X-ray correlation during this period.
We started our observations just before the first  X-ray maximum, and the radio demonstrated a smooth decay until 11 days before periastron. The rapid radio increase after periastron (at $\approx+14$  days) occurred with the rise of the X-ray emission, and  closely matched 
the data from previous periastron passages. This time, however 
the X-ray/radio correlation disappeared right after the beginning of the X-ray flux decay $\sim 20$ days after the periastron, contrary to the behaviour observed during the 2021 periastron passage, where the correlation was seen until  $\sim 30$ days after periastron.

The absence of a correlation can be connected to different populations of electrons producing radio and X-ray emission as suggested by~\citet{Chernyakova21_psrb}. 
X-ray light curve allows 
two interpretations of the late-phases of the second X-ray peak: \textit{(i)}: a broad 2nd peak with occasional initial radio/X-ray coupling; \textit{(ii)}: the presence of the 3rd X-ray peak, similar to one observed in 2021~\citep{Chernyakova20_psrb}, but appearing much earlier  at $\sim +20$~days.

The radio flux density in 2024 reached a historical maximum $\approx$16 days after periastron \citep[see e.g. compilations of historical radio data in][and references therein]{Chernyakova21_psrb}. This maximum was followed by a rapid decay during which the radio emission dropped to a level comparable with the level of the pulsed emission at $\sim30$ days after the periastron, much earlier than previously observed. The indexes of the radio emission  are presented in the fourth panel of Figure \ref{fig:multiwavelength_lc} (green circles) along with the X-ray indexes (blue points) and the slope between the radio to X-ray (magenta squares). In 2024 the indexes of the radio data were much more chaotic than in 2021, where they were systematically softer than the X-ray ones.

The GeV-band emission demonstrated a sharp rise  from $\sim+27$\,days, which peaked at $\sim +30$~days, at a time comparable to the 2010 periastron and substantially earlier than observed for the 2021 periastron. Such an early rise of the GeV emission contradicts the precession disk model proposed by \cite{B1259-precession}  After a gradual decay of the emission a second rise was observed at $\sim 60$ to $+74$~days. A number of short down to 6~min duration) flares with energetics exceeding the spin-down luminosity up to a factor of $\sim 5$ were observed during both periods of high GeV flux, see Fig.~\ref{fig:fermi_lc}.

The interaction of the pulsar wind with the outflow of  LS2883 led to the formation of a shock cone directed away from the massive star,  with the axis lying along the line connecting the pulsar and LS 2883 \citep[see Figure 5 in][]{Chernyakova2017}).  In the model of~\citet{Chernyakova2017} the GeV emission is explained as emission of the relativistic pulsar wind electrons flowing along the shock with a moderate Lorentz factor $\gamma\sim 3$. Similar to the hollow cone model described in \citet{2008ApJ...672L.123N} the GeV \lat flare occurs when the pulsar wind/decretion disk interaction cone is oriented towards the observer. 
The short $\sim 15$~min-long sub-flares with energetics significantly exceeding the spin-down luminosity are explained as bremsstrahlung radiation that occurs due to the entrance of the clumps of the decretion disk into the interaction cone. 

In 2024 the \lat flare occurred close to the end of the second X-ray peak, during the period of quasi-random X-ray variability. As discussed above, during this period the regular structure of the decretion disk was strongly perturbed by the pulsar. Still, the density of the material was sufficiently high to form a relatively thin emission cone and the rise of the GeV emission at $+20$ to $+37$~days.

The second rise of the GeV emission ($+60$ to $+74$~days)  can be connected to the eventual disruption of the disk and the production of a large number of clumps which reach the interaction cone. This hypothesis is further supported by a larger number of short sub-flares seen during the second rise of the GeV emission, comparable to the first one (see Figure \ref{fig:fermi_lc}).
The Bayesian-blocks time binning light curve indicates the presence of 5 sub-flares with the mean flux exceeding the spin-down luminosity for the period $+27$ to $+37$~days and 7 sub-flares for the period $+60$ to $+74$~days. Such a behavior can be indicative of a larger number of clumps reaching the interaction cone when the pulsar was significantly far from periastron.

Interestingly, the Bayesian analysis revealed the presence of short timescale variability well before periastron at $\sim -54$ and $\sim -20$~days, with fluxes comparable to the spin-down luminosity. While the \lat sensitivity does not allow us to firmly conclude on the exact flux level of these flaring events we note that pre-periastron flares with spin-down-exceeding luminosity are hard to interpret within the discussed model. We also note that pre-periastron flares with a similar luminosity were not observed during the 2010-2021 periastron passages.

The spectral behavior of the system in the GeV band differs from the one observed during the 2017-2021 periastron passages. Contrary to previous periastra, in 2024, \psrb is detected with a higher significance before periastron ($-20$ to $0$~days, $TS\sim 17$) and a lower significance after periastron ($0$ to $+20$~days, $TS\sim 6.5$). The low significance of the detection during these time periods did not allow us to detect a spectral curvature contrary to previous passages. The flaring-period-averaged flux ($+19$ to $+77$~days) is also somewhat lower than the one seen in 2017 (see Fig.~\ref{fig:fermi_spectra} and Table~\ref{tab:fermi_bestfits}). 

While detailed modeling of the \psrb emission around 2024 periastron passage is beyond the scope of this work, we argue, that the observed changes in the GeV band spectrum can be connected to a difference in the particles' acceleration efficiency in the presence of a denser/larger disk and/or by modified escape and cooling times.

\section*{Acknowledgments}
 The authors acknowledge support by the state of Baden-W\"urttemberg through~bwHPC.
 We thank Rajan Chhetri and Jamie Stevens for their assistance during the ATCA observations.
 This work made use of data supplied by the UK Swift Science Data Centre at the University of Leicester. The Australia Telescope Compact Array (ATCA) is part of the Australia Telescope National Facility (\url{https://ror.org/05qajvd42}) which is funded by the Australian Government for operation as a National Facility managed by CSIRO. We acknowledge the Gomeroi people as the Traditional Owners of the ATCA observatory site. Some of the observations reported in this paper were obtained with the Southern African Large Telescope (SALT) under program 2024-1-SCI-040 (PI: B. van Soelen). The authors thank the European Space Agency (ESA) for the support in the framework of the PRODEX Programme, PEA 4000120711. 
 The authors wish to acknowledge financial support from the Centre for Astrophysics and Relativity at DCU.

\section*{Data Availability}
The data underlying this article will be shared on reasonable request to the corresponding authors.

\def\aj{AJ}%
\def\actaa{Acta Astron.}%
\def\araa{ARA\&A}%
\def\apj{ApJ}%
\def\apjl{ApJ}%
\def\apjs{ApJS}%
\def\ao{Appl.~Opt.}%
\def\apss{Ap\&SS}%
\def\aap{A\&A}%
\def\aapr{A\&A~Rev.}%
\def\aaps{A\&AS}%
\def\azh{AZh}%
\def\baas{BAAS}%
\def\bac{Bull. astr. Inst. Czechosl.}%
\def\caa{Chinese Astron. Astrophys.}%
\def\cjaa{Chinese J. Astron. Astrophys.}%
\def\icarus{Icarus}%
\def\jcap{J. Cosmology Astropart. Phys.}%
\def\jrasc{JRASC}%
\def\mnras{MNRAS}%
\def\memras{MmRAS}%
\def\na{New A}%
\def\nar{New A Rev.}%
\def\pasa{PASA}%
\def\pra{Phys.~Rev.~A}%
\def\prb{Phys.~Rev.~B}%
\def\prc{Phys.~Rev.~C}%
\def\prd{Phys.~Rev.~D}%
\def\pre{Phys.~Rev.~E}%
\def\prl{Phys.~Rev.~Lett.}%
\def\pasp{PASP}%
\def\pasj{PASJ}%
\def\qjras{QJRAS}%
\def\rmxaa{Rev. Mexicana Astron. Astrofis.}%
\def\skytel{S\&T}%
\def\solphys{Sol.~Phys.}%
\def\sovast{Soviet~Ast.}%
\def\ssr{Space~Sci.~Rev.}%
\def\zap{ZAp}%
\def\nat{Nature}%
\def\iaucirc{IAU~Circ.}%
\def\aplett{Astrophys.~Lett.}%
\def\apspr{Astrophys.~Space~Phys.~Res.}%
\def\bain{Bull.~Astron.~Inst.~Netherlands}%
\def\fcp{Fund.~Cosmic~Phys.}%
\def\gca{Geochim.~Cosmochim.~Acta}%
\def\grl{Geophys.~Res.~Lett.}%
\def\jcp{J.~Chem.~Phys.}%
\def\jgr{J.~Geophys.~Res.}%
\def\jqsrt{J.~Quant.~Spec.~Radiat.~Transf.}%
\def\memsai{Mem.~Soc.~Astron.~Italiana}%
\def\nphysa{Nucl.~Phys.~A}%
\def\physrep{Phys.~Rep.}%
\def\physscr{Phys.~Scr}%
\def\planss{Planet.~Space~Sci.}%
\def\procspie{Proc.~SPIE}%
\let\astap=\aap
\let\apjlett=\apjl
\let\apjsupp=\apjs
\let\applopt=\ao
\bibliography{references.bib}
\end{document}